\documentclass[prd,nofootinbib,onecolumn,preprint,11pt]{revtex4}

\usepackage{graphicx}
\usepackage{bm}
\usepackage{float}
\usepackage{subfigure}
\usepackage{amsmath}
\usepackage{amsfonts}
\usepackage{color}
\usepackage{slashed}
\usepackage{fancyhdr}

\newcommand{\gb}{\bar{g}}

\newcommand{\sbar}{\bar{\sigma}}

\newcommand{\dd}{\text{d}}

\newcommand{\D}{\mathcal{D}}

\newcommand{\mma}{\textsf{Mathematica}}

\begin{document}

\title{A Line Source In Minkowski For The de Sitter Spacetime Scalar Green's Function: Massive Case}

\author{Yi-Zen Chu}
\affiliation{
Center for Particle Cosmology, Department of Physics and Astronomy, University of Pennsylvania, Philadelphia, Pennsylvania 19104, USA
}
%\date{}

\begin{abstract}
\noindent For certain classes of space(time)s embeddable in a higher dimensional flat space(time), it appears possible to compute the minimally coupled massless scalar Green's function in the former by convolving its cousin in the latter with an appropriate scalar charge density. The physical interpretation is that beings residing in the higher dimensional flat space(time) may set up sources to fool the observer confined on the lower dimensional curved submanifold that she is detecting the field generated by a space(time) point source in her own world. In this paper we extend the general formula to include a non-zero mass. We then employ it to derive the Green's function of the massive wave operator in $(d \geq 2)$-dimensional de Sitter spacetime and that of the Helmholtz differential operator -- the Laplacian plus a ``mass term" -- on the $(d \geq 2)$-sphere. For both cases, the trajectories of the scalar sources are the same as that of the massless case, while the required scalar charge densities are determined by solving an eigenvalue equation. To source these massive Green's functions, we show that the $(d+1)$-dimensional Minkowski/Euclidean experimentalists may choose to use either massive or massless scalar line charges. In de Sitter spacetime, the embedding method employed here leads directly to a manifest separation between the null cone versus tail terms of the Green's functions.
\end{abstract}

\maketitle

\section{Introduction}

In space(time)s enjoying a high degree of symmetry, the Green's function $G[x,x']$ of its associated wave operator $\mathcal{W}$ for a given field theory can oftentimes be solved by the separation-of-variables technique. The solution obtained in this manner is usually an infinite sum (and/or integral(s)) involving the orthonormal mode functions $\{\langle x | \lambda \rangle \}$ obeying the eigenvector equation\footnote{We will employ Dirac's bra-ket notation throughout this paper.}
\begin{align}
\mathcal{W}_x \langle x | \lambda \rangle = \lambda \langle x | \lambda \rangle .
\end{align}
The mode decomposition reads
\begin{align}
\label{GreensFunction_ModeExpansion_Schematic}
G[x,x'] = \sum_\lambda \frac{\langle x | \lambda \rangle \langle \lambda | x' \rangle}{\lambda} .
\end{align}
Such an expansion may not be the most useful, unfortunately, if one is seeking to understand the full causal structure of signals produced by physical sources moving about in the curved spacetime of interest. In generic curved spacetimes, unlike in 4D Minkowski spacetime, even massless fields do not propagate strictly on the null cone. (The portion of the field propagating inside the light cone of its source is known in the literature as the ``tail".) When the observer at $x$ can be linked to the source at $x'$ via a unique geodesic, for instance, the 4D Green's function $G[x,x']$ ought to be expressible as the sum of a term describing scalar waves propagating on the null cone of $x'$ and a separate term describing tail propagation -- see, for e.g., eq. (14.4) of \cite{Poisson:2011nh}. However, this property is not reflected very well by the expansion in eq. \eqref{GreensFunction_ModeExpansion_Schematic}. Furthermore, already in the weak field limit of Schwarzschild/Kerr spacetime, the behavior of tails changes abruptly, going from the region $t-t' < r+r'$ to the region $t-t' > r+r'$ \cite{DeWittDeWitt:1964,PfenningPoisson:2000zf,Chu:2011ip}. (Here, $t-t'$ is the elapsed time between observation and emission; and $r$ and $r'$ are, respectively, the radial coordinates of the observer and source.) This means the tail effect most likely cannot be captured comprehensively by merely pushing to very higher orders in perturbation theory methods that can be utilized to compute the tail part of the Green's functions when $x$ and $x'$ are close to one another.\footnote{To be sure, the existence of the sharp boundary at $t-t'=r+r'$ in \cite{DeWittDeWitt:1964,PfenningPoisson:2000zf,Chu:2011ip} is because the central source is a point mass. Any realistic finite size material body would likely yield a smeared boundary centered at $t-t'=r+r'$, but the behavior of the tail effect on either side would still be quite different.} It is for these reasons that one may be motivated to search for alternate means to calculate Green's functions in curved space(time)s.

Understanding the tail effect in black hole spacetimes, in particular, is important for the radiation reaction aspect of modeling the gravitational waves generated by compact bodies orbiting super-massive black holes. The tail portion of the gravitational field generated by a compact body at some spacetime point in the past necessarily has a non-zero impact on its motion in the future -- see Fig. (2) of \cite{Poisson:2011nh} for a visual illustration of this ``self-force".\footnote{A sample of recent work on the scalar Green's function in Schwarzschild spacetime can be found in \cite{Casals:2013mpa,Casals:2012ng,Zenginoglu:2012xe}} The tail effect also imparts a signature on the gravitational wave signal produced by in-spiraling comparable mass compact neutron star/black hole binaries, a major class of sources for detectors like advanced LIGO; for, the gravitational radiation they produce can backscatter off the spacetime curvature of the binary system itself, and lead to a modulation in the frequency spectrum of the waveforms \cite{Blanchet:1992br, Blanchet:1993ec}. Moreover, if gravitational and electromagnetic waves traveled over cosmological distances before reaching Earth, they would also develop tails due to their interaction with the curved spacetime of the expanding universe at large. (Due to the conformal invariance of the Maxwell action, photons develop a tail iff the universe is not perfectly isotropic and homogeneous \cite{Chu:2011ip}.) For these physical reasons, the search for different ways to calculate curved spacetime Green's functions is not merely an academic exercise in mathematical physics.

Recently, in \cite{Chu:2013hra}, we have wondered whether Green's functions in cosmological and black hole spacetimes can be solved by exploiting the known perspective \cite{Robertson:1933zz,Kasner,Fronsdal:1959zza} that the general Friedmann-Lema\^{i}tre-Robertson-Walker universe and Schwarzschild black hole can be described as curved surfaces embedded in, respectively, 5 and 6 dimensional Minkowski spacetime. In physical terms, we also asked, for a given curved spacetime embedded in some higher dimensional Minkowski, if it were always possible for a family of beings living in the latter to arrange appropriate source(s) so that the observer confined on the curved sub-surface would interpret the resulting field as due to that of a spacetime point source in her own world. The answer turns out to be in the affirmative for $(d \geq 2)$-dimensional de Sitter spacetime, which can be viewed as a hyperboloid in $(d+1)$-dimensional Minkowski. Using the massless scalar Green's function in Minkowski, we computed its counterpart in de Sitter by first identifying the source to be a straight line emanating from the Minkowski origin, piercing the de Sitter hyperboloid orthogonally at $x'$, and extending to infinity. Following that, we solved for the appropriate scalar charge density, which is related to the null vectors of the ``wave operator" perpendicular to the de Sitter hyperboloid. This embedding method lead directly to an explicit separation between the null cone versus tail pieces of the de Sitter Green's function in arbitrary dimensions. Furthermore, the intermediate steps revealed that the source of the light cone part of the scalar signal -- as seen by the de Sitter observer at $x$ -- can be identified by the higher dimensional Minkowski experimentalist to be the point $x'$ where the line intersects the de Sitter hyperboloid; while the de Sitter tail portion of the scalar signal is sourced by the rest of the infinite line lying within the Minkowski null cone of $x$. This insight is very similar in spirit to the embedding explanation of the massless tail effect in odd dimensional Minkowski spacetime \cite{SoodakTiersten}, by viewing it as a surface embedded in one higher (and hence even) dimensional Minkowski, where there are no tails.

In this paper, we shall extend the results of \cite{Chu:2013hra} to include a non-zero mass $m$: we will solve the Green's function of the massive scalar wave operator $\Box + m^2$ in de Sitter and on the $d$-sphere. In section \eqref{Section_Generalities} we lay down the general formulas that would allow us to calculate the massive Green's function in certain classes of curved space(time)s using its counterpart in the ambient higher dimensional flat space(time)s. In section \eqref{Section_Flat} we discuss the massive Green's functions in flat Minkowski and Euclidean space(time)s for all relevant dimensions. In section \eqref{Section_deSitter} we apply the general formulas to work out explicit expressions for the retarded and advanced massive Green's functions in $(d \geq 2)$-dimensional de Sitter spacetime. In section \eqref{Section_dSphere} we solve the Green's function of the Helmholtz operator on the $(d\geq 2)$-sphere. This is a follow-up of the calculation in \cite{Chu:2013hra}, where we studied how our general formulas broke down due to topology; here, there is no topological obstruction to the solution of a massive field sourced by a single point charge on the $d$-sphere. Finally, in section \eqref{Section_End}, we summarize our findings.

\section{Generalities}
\label{Section_Generalities}

We will suppose that the $d$ dimensional curved spacetime $g_{\mu\nu}[x]$ of interest is embeddable in some $(d+n)$ dimensional flat Minkowski spacetime,
\begin{align}
\label{MinkowskiMetric}
\dd s^2 						&= \eta_{\mathfrak{A}\mathfrak{B}} \dd X^{\mathfrak{A}} \dd X^{\mathfrak{B}} \equiv \dd X \cdot \dd X , \\
\eta_{\mathfrak{A}\mathfrak{B}} &\equiv \text{diag}[1,-1,\dots,-1],
\end{align}
with Cartesian coordinates $\{X^\mathfrak{A} | \mathfrak{A} = 0,1,2,\dots,d+n-1 \}$. By this we mean it is possible to perform a coordinate transformation
\begin{align}
X^{\mathfrak{A}} 
	&\to X^{\mathfrak{A}}[x,y] \\
\dd X^{\mathfrak{A}} 
	&\to \frac{\partial X^{\mathfrak{A}}[x,y]}{\partial x^\alpha} \dd x^\alpha + \frac{\partial X^{\mathfrak{A}}[x,y]}{\partial y^\text{B}} \dd y^\text{B},
\end{align}
using the coordinates $\{x^\mu,y^\text{A}|\mu=0,1,2,\dots,d-1; \ \text{A} = 1,2,3,\dots,n \}$, so that the flat metric in \eqref{MinkowskiMetric} now becomes
\begin{align}
\label{Metric_GeneralEmbeddingForm}
\dd s^2 &= P^2[y] \bar{g}_{\mu\nu}[x] \dd x^\mu \dd x^\nu + g^\perp_\text{AB}[y] \dd y^\text{A} \dd y^\text{B} ,
\end{align}
such that $g_{\mu\nu}[x]$ is realized on some surface $y=y_0$, i.e.
\begin{align}
\label{Metric_OfActualInterest}
g_{\mu\nu}[x] = P^2[y=y_0] \bar{g}_{\mu\nu}[x] \equiv P_0^2 \bar{g}_{\mu\nu}[x] .
\end{align}
Here, $\bar{g}_{\mu\nu}$ is independent of $y$ while $P$ and $g^\perp_\text{AB}$ are independent of $x$. The massive wave operator $\Box + m^2$ (of $g_{\mu\nu}[x]$) acting on a scalar $\psi$ reads
\begin{align}
\label{ScalarWaveOperator_gmunu}
\left(\Box + m^2\right)\psi 
= \frac{1}{P_0^2 \sqrt{|\gb|}} \partial_\mu \left( \sqrt{|\gb|} \gb^{\mu\nu} \partial_\nu \psi \right) + m^2 \psi
\equiv \frac{1}{P_0^2} \left(\overline{\Box} + P_0^2 m^2 \right)\psi ,
\end{align}
with $\sqrt{|\gb|}$ denoting the square root of the absolute value of the determinant of $\gb_{\mu\nu}$; and $\gb^{\mu\nu}$ is its inverse. 

The central assertion in this paper is that the Green's function $G_d[x,x']$ of $\Box + m^2$, obeying the equation
\begin{align}
\label{GreensFunction_DefiningEquation}
\left(\Box_x + m^2\right)G_d[x,x'] = \left(\Box_{x'} + m^2\right) G_d[x,x'] 
= \frac{\delta^{(d)}[x-x']}{\left\vert g[x] g[x'] \right\vert^{1/4}} 
= \frac{\delta^{(d)}[x-x']}{P^d[y_0] \left\vert \gb[x] \gb[x'] \right\vert^{1/4}} 
\end{align}
can be gotten from its $(d+n)$-dimensional Minkowski relative $\overline{G}_{d+n}[X-X']$, through the general formula
\begin{align}
\label{GreensFunction_GeneralFormula}
G_d[x,x'] 
= \int \dd^n y' \sqrt{|g^\perp[y']|} \frac{P^{d-2}[y']}{P^{d-2}[y_0]} 
\frac{\langle y' | P_0^2 m^2 \rangle}{\langle y_0 | P_0^2 m^2 \rangle} 
\overline{G}_{d+n}\Big[ X[x,y_0] - X'[x',y'] \Big] .
\end{align}
The $\langle y' | P_0^2 m^2 \rangle$ is any one of the eigenvector(s) of $\mathcal{D} + P^2 m^2$ with eigenvalue $P_0^2 m^2$,
\begin{align}
\label{EigenvectorEquation}
\left(\D_y + P^2[y] m^2\right) \langle y | P_0^2 m^2 \rangle
&\equiv \frac{\partial_\text{A} \left( P^d \sqrt{|g^\perp|} (g^\perp)^\text{AB} \partial_\text{B} \langle y | P_0^2 m^2 \rangle \right)}{P^{d-2} \sqrt{|g^\perp|}} + P^2[y]m^2 \langle y | P_0^2 m^2 \rangle \nonumber\\
&= P_0^2 m^2 \langle y | P_0^2 m^2 \rangle .
\end{align}
The $|g^\perp[y']|$ in eq. \eqref{GreensFunction_GeneralFormula} is the absolute value of the determinant of the metric $g^\perp_\text{AB}[y']$. The $\overline{G}_{d+n}[X-X']$ themselves, which can be found in equations \eqref{GreensFunction_Minkowski_RetardedAdvanced} through \eqref{GreensFunction_Minkowski_Odd} below, obey the equations
\begin{align}
\label{GreensFunction_Minkowski_CartesianWaveEquation}
\left(\eta^{\mathfrak{A}\mathfrak{B}} \partial_{\mathfrak{A}}\partial_{\mathfrak{B}} + m^2\right)\overline{G}_{d+n}[X-X']
= \left(\eta^{\mathfrak{A}\mathfrak{B}} \partial_{\mathfrak{A}'}\partial_{\mathfrak{B}'} + m^2\right)\overline{G}_{d+n}[X-X'] 
= \delta^{(d+n)}[X-X'],
\end{align}
with the unprimed indices denoting derivatives with respect to $X$ and the primed indices derivatives with respect to $X'$. Note that, in terms of the coordinates $x$ and $y$, the massive flat spacetime wave operator becomes
\begin{align}
\left(\eta^{\mathfrak{A}\mathfrak{B}} \partial_\mathfrak{A} \partial_\mathfrak{B} + m^2\right)\psi 
= \frac{1}{P^2[y]} \left( \overline{\Box}_x + \D_y + P^2[y]m^2 \right) \psi .
\end{align}
Therefore, multiplying both sides of the flat spacetime Green's function equation \eqref{GreensFunction_Minkowski_CartesianWaveEquation} by $P^2$ and translating it into $x$ and $y$ coordinates hands us
\begin{align}
\label{GreensFunction_Minkowski_CurvedCoordinates}
\left( \overline{\Box}_x + \D_y + P^2[y]m^2 \right) \overline{G}_{d+n}[X-X']
&= \left( \overline{\Box}_{x'} + \D_{y'} + P^2[y']m^2 \right) \overline{G}_{d+n}[X-X'] \nonumber\\
&= \frac{\delta^{(d)}[x-x']}{P^{\frac{d-2}{2}}[y] P^{\frac{d-2}{2}}[y'] \sqrt[4]{|\gb[x]\gb[x']|}} \frac{\delta^{(n)}[y-y']}{\sqrt[4]{|g^\perp[y] g^\perp[y']|}} ,
\end{align}

{\bf Justification of eq. \eqref{GreensFunction_GeneralFormula}} \qquad Following arguments in \cite{Chu:2013hra}, we consider applying $\overline{\Box}_{x'} + P_0^2 m^2$ on both sides of the general formula in eq. \eqref{GreensFunction_GeneralFormula}. After inter-changing the order of integration and differentiation, re-writing 
\begin{align}
\label{ScalarWaveOperator_SeparatedForm}
\left(\Box_{x'} + m^2 \right)\psi &= \frac{1}{P_0^2} \left(\overline{\Box}_{x'} + P_0^2 m^2\right)\psi \nonumber\\
&= \frac{P^2[y']}{P_0^2} \left(\eta^{\mathfrak{A'}\mathfrak{B'}} \partial_{\mathfrak{A}'} \partial_{\mathfrak{B}'} + m^2\right) \psi 
	- \frac{1}{P_0^2} \left\{ \D_{y'} + m^2 (P^2[y'] - P_0^2) \right\}\psi ,
\end{align}
and exploiting the wave equation of \eqref{GreensFunction_Minkowski_CartesianWaveEquation}, we may deduce that
\begin{align}
\label{GreensFunction_GeneralFormula_Justification_II}
&\left(\Box_{x'} + m^2 \right)G_d[x,x'] 
= \frac{\delta^{(d)}[x-x']}{P^d[y_0] \sqrt[4]{\gb[x] \gb[x']}} \\
&\qquad\qquad
- \int \dd^n y' \sqrt{|g^\perp[y']|} \frac{P^{d-2}[y']}{P^d[y_0]} 
\frac{\left\{ \D_{y'} + m^2 (P^2[y'] - P_0^2) \right\} \langle y' | P_0^2 m^2 \rangle}{\langle y_0 | P_0^2 m^2 \rangle} 
\overline{G}_{d+n}\Big[ X[x,y_0] - X'[x',y'] \Big] \nonumber
\end{align}
In eq. \eqref{GreensFunction_GeneralFormula_Justification_II}, we have integrated-by-parts the operator $\D_{y'}$ and shifted it to act on $\langle y' | P_0^2 m^2 \rangle$. (That the boundary terms are zero can be checked explicitly for the main results in this paper.) Also, in eq. \eqref{GreensFunction_GeneralFormula_Justification_II}, we see the reason for the $P^{d-2}[y_0] \langle y_0 | P_0^2 m^2 \rangle$ in the denominator of eq. \eqref{GreensFunction_GeneralFormula}: this ensures the measure multiplying the $\delta$-functions on the right hand side is the desired one. From the second line of eq. \eqref{GreensFunction_GeneralFormula_Justification_II}, we may now verify that the Green's function equation in eq. \eqref{GreensFunction_DefiningEquation} is satisfied if  $\langle y' | P_0^2 m^2 \rangle$ obeys the eigenvalue equation \eqref{EigenvectorEquation}.

For the de Sitter and $d$-sphere calculations below, it will be useful to recognize that the $\langle y_0 | P_0^2 m^2 \rangle$ in the denominator of the general formula in eq. \eqref{GreensFunction_GeneralFormula} will cancel out in the final answer, upon the evaluation of the $\int \dd^n y'$ integral(s). We give a brief (heuristic) argument here, by assuming that the Green's function admits a mode sum expansion and that a separation-of-variables ansatz holds for the mode functions, namely
\begin{align}
\label{GreensFunction_Minkowski_ModeExpansion}
\overline{G}_{d+n}\Big[ X[x,y_0]-X'[x',y'] \Big]
= \sum_{\lambda,\lambda^\perp} \frac{\langle x | \lambda \rangle \langle y_0 | \lambda^\perp \rangle \cdot \langle \lambda^\perp | y' \rangle \langle \lambda | x' \rangle}{\lambda+\lambda^\perp} ,
\end{align}
where $\langle x | \lambda \rangle$ is the orthonormal eigenvector of $\overline{\Box}_x$ with eigenvalue $\lambda$, and $\langle y | \lambda^\perp \rangle$ is the orthonormal eigenvector of $\D_y + (P[y] \cdot m)^2$ with eigenvalue $\lambda^\perp$. (Refer to eq. \eqref{GreensFunction_Minkowski_CurvedCoordinates}.) Because $\D_{y'} + (P[y'] \cdot m)^2$ is hermitian with respect to the integration measure $\sqrt{|g^\perp[y']|} P^{d-2}[y']$, we may view the integration $\int \dd^n y'$ as a projection of $\overline{G}_{d+n}$ along the orthonormal eigenvector $\langle y' | \lambda^\perp = P_0^2 m^2 \rangle$. This collapses the $\lambda^\perp$-sum in eq. \eqref{GreensFunction_Minkowski_ModeExpansion}, leaving $\langle y_0 | \lambda^\perp = P_0^2 m^2 \rangle \cdot \langle x | \lambda \rangle \langle \lambda | x' \rangle$ in the numerator; the presence of $\langle y_0 | \lambda^\perp = P_0^2 m^2 \rangle$ then confirms our assertion.

We further observe that any one of the eigenvectors of $\D + (P \cdot m)^2$ with eigenvalue $P_0^2 m^2$ ought to lead to the same result in eq. \eqref{GreensFunction_GeneralFormula}, indicating there is more than one scalar charge density in the ambient Minkowski that can mimic the same point charge in the curved world of the observer. (For example, in curved spacetimes embeddable in Minkowski of one higher dimension, the eigenvector equation \eqref{EigenvectorEquation} reduces to a second order ordinary differential equation, and therefore admits two distinct scalar charge densities.) Notice, furthermore, that the trajectory of the scalar source that the higher dimensional experimentalist has set up in the general formula of eq. \eqref{GreensFunction_GeneralFormula} is the same as the massless case. What appears to be different is the use of the massive flat spacetime Green's function, and the scalar charge density described by the eigenvector $\langle y' | P_0^2 m^2 \rangle$. As we shall witness in the de Sitter and $d$-sphere computations below, however, the analog of eq. \eqref{GreensFunction_GeneralFormula} will reduce to one involving the massless Green's function in flat space(time).

In this section, analogous statements hold true if we replaced all instances of ``Minkowski" and ``flat spacetime" with ``Euclidean space," and $\eta_{\mathfrak{A}\mathfrak{B}}$ with $\delta_{\mathfrak{A}\mathfrak{B}}$ (the Kronecker delta). The corresponding massive Euclidean Green's function can be found in equations \eqref{GreensFunction_Euclidean_Even} and \eqref{GreensFunction_Euclidean_Odd}. The primary physical difference between the Euclidean versus Lorentzian setups is that there is no issue of causal influence in the former but for the latter, one may ask, if cause precedes effect in the ambient Minkowski, does that necessarily imply that cause precedes effect on the curved sub-manifold? (The answer is, yes, for the closed slicing representation of de Sitter spacetime.)

\section{Flat Space(time)s}
\label{Section_Flat}

In this section we record the massive Green's functions in flat space(time).

{\bf Minkowski Spacetime} \qquad For $d$ dimensional Minkowski spacetime and for real $m$, the retarded $\overline{G}_d^+$ and advanced $\overline{G}_d^-$ Green's functions, obeying eq. \eqref{GreensFunction_Minkowski_CartesianWaveEquation} are given by
\begin{align}
\label{GreensFunction_Minkowski_RetardedAdvanced}
\overline{G}^\pm_d[X-X'] = \Theta[\pm(X^0-X'^0)] \overline{\mathcal{G}}_d[\sbar] .
\end{align}
For even $d \geq 2$, the symmetric Green's function $\overline{\mathcal{G}}_d$ is
\begin{align}
\label{GreensFunction_Minkowski_Even}
\overline{\mathcal{G}}_d[\sbar] 
= \frac{1}{2(2\pi)^{\frac{d-2}{2}}} \left( \frac{\partial}{\partial \bar{\sigma}} \right)^{\frac{d-2}{2}}
\left(\Theta[\bar{\sigma}] J_0\left[m\sqrt{2\bar{\sigma}}\right]\right), 
\end{align}
(the $J_0$ is the Bessel function of the first kind) while for odd $d \geq 3$, it becomes instead
\begin{align}
\label{GreensFunction_Minkowski_Odd}
\overline{\mathcal{G}}_d[\sbar] 
= \frac{1}{(2\pi)^{\frac{d-1}{2}}} \left( \frac{\partial}{\partial \bar{\sigma}} \right)^{\frac{d-3}{2}} \left( \Theta\left[ \bar{\sigma} \right] \frac{\cos\left[ m \sqrt{2 \bar{\sigma}} \right]}{\sqrt{2 \bar{\sigma}}} \right) .
\end{align}
We have written these results in terms of Synge's world function $\bar{\sigma}$, which is half the square of the geodesic distance between the observer at $X$ and the emitter at $X'$. It is
\begin{align}
\label{Minkowski_SyngeWorldFunction}
\bar{\sigma} \equiv \frac{1}{2} \left(X-X'\right)^2 
\equiv \frac{1}{2} \eta_{\mathfrak{A}\mathfrak{B}} \left(X-X'\right)^\mathfrak{A} \left(X-X'\right)^\mathfrak{B} .
\end{align}
The retarded or advanced conditions are encoded using the step function
\begin{align}
\Theta[z] 
&= 1, \qquad z \geq 0 \nonumber\\
&= 0, \qquad z < 0 .
\end{align}

{\bf Euclidean Space} \qquad In Euclidean space the Green's function of the massive Laplace operator satisfies
\begin{align*}
\left( -\delta^{\mathfrak{A}\mathfrak{B}} \partial_\mathfrak{A} \partial_\mathfrak{B} + m^2 \right) \overline{G}^\text{(E)}_d[\vec{X}-\vec{X}'] 
= \left( -\delta^{\mathfrak{A}\mathfrak{B}} \partial_{\mathfrak{A}'} \partial_{\mathfrak{B}'} + m^2 \right) \overline{G}^\text{(E)}_d[\vec{X}-\vec{X}']
= \delta^{(d)}[\vec{X}-\vec{X}'] .
\end{align*}
If $m$ is real, this equation is the static limit of the massive wave equation: $\partial_t^2 -\delta^{\mathfrak{A}\mathfrak{B}} \partial_\mathfrak{A} \partial_\mathfrak{B} + m^2 \to -\delta^{\mathfrak{A}\mathfrak{B}} \partial_\mathfrak{A} \partial_\mathfrak{B} + m^2$.\footnote{It is for this reason that we have chosen to put a negative sign in front of the Laplacian. The $m \to 0$ limit of the results in this paper, for example equations \eqref{GreensFunction_Euclidean_Even} and \eqref{GreensFunction_Euclidean_Odd}, are therefore negative of the corresponding results in \cite{Chu:2013hra}.} If $m$ is purely imaginary, this equation can be viewed as the massless wave operator written in frequency space, namely $\partial_t^2 -\delta^{\mathfrak{A}\mathfrak{B}} \partial_\mathfrak{A} \partial_\mathfrak{B} \to -\omega^2 -\delta^{\mathfrak{A}\mathfrak{B}} \partial_\mathfrak{A} \partial_\mathfrak{B}$, with $m \equiv i\omega$ and $\omega \in \mathbb{R}$.

Denote the Euclidean distance between observer and source by
\begin{align}
\bar{R} \equiv \left\vert \vec{X}-\vec{X}' \right\vert .
\end{align}
For even $d \geq 2$, the massive Euclidean Green's function reads
\begin{align}
\label{GreensFunction_Euclidean_Even}
\overline{G}^\text{(E)}_{\text{even }d}[\vec{X}-\vec{X}']
= \frac{1}{2 \pi} \left(-\frac{1}{2\pi \bar{R}} \frac{\partial}{\partial \bar{R}}\right)^{\frac{d-2}{2}} K_0[m \bar{R}], 
\end{align}
($K_0$ is the modified Bessel function.) For odd $d \geq 1$, it is
\begin{align}
\label{GreensFunction_Euclidean_Odd}
\overline{G}^\text{(E)}_{\text{odd }d}[\vec{X}-\vec{X}']
= \left(-\frac{1}{2\pi \bar{R}} \frac{\partial}{\partial \bar{R}}\right)^{\frac{d-1}{2}} \frac{e^{-m\bar{R}}}{2 m} .
\end{align}
{\bf Recursion Relations} \qquad As discussed in \cite{Chu:2013hra,SoodakTiersten}, we may view $d$-dimensional space(time) as being a surface embedded in $(d+1)$ dimensional space(time). For Minkowski spacetime, we may identify
\begin{align}
P = 1, \qquad g^\perp_\text{AB} \dd y^\text{A} \dd y^\text{B} = -(\dd X^d)^2
\end{align}
so that the eigenvector equation \eqref{EigenvectorEquation} is
\begin{align}
-\frac{\dd^2 \langle X^d \vert m^2 \rangle}{\dd (X^d)^2} + m^2 \langle X^d \vert m^2 \rangle = m^2 \langle X^d \vert m^2 \rangle .
\end{align}
The two independent solutions are a constant and $X^d$. 

If we choose $\langle X^d \vert m^2 \rangle = $ constant as the charge density, \cite{Chu:2013hra} has already proven that the Minkowski Green's functions obey a recursion relation
\begin{align}
\overline{G}^\pm_{d+2}[X-X'] = \frac{1}{2\pi} \frac{\partial}{\partial \sbar} \overline{G}^\pm_d[X-X']
\end{align}
if we agree not to differentiate the $\Theta[\pm(X^0-X'^0)]$. After carrying out the same line of reasoning for the Euclidean space counterparts, we will find that they obey
\begin{align}
\overline{G}^\text{(E)}_{d+2}[\vec{X}-\vec{X}'] = -\frac{1}{2\pi \overline{R}} \frac{\partial}{\partial \overline{R}} \overline{G}^\text{(E)}_d[\vec{X}-\vec{X}'] .
\end{align}
(The validity of these recursion relations do not depend on the value of $m$.) On a practical level, to work out the $d$ dimensional Minkowski and Euclidean Green's functions, it suffices to evaluate the integrals
\begin{align}
\overline{G}^\pm_d[\vec{X}-\vec{X}'] = \int_\pm \frac{\dd^d k}{(2\pi)^d} \frac{e^{-ik \cdot (X-X')}}{-k^2+m^2}
\end{align}
for $d=2,3$ (the $\int_\pm$ represent the retarded and advanced contour prescriptions for the $k_0$ integral), and
\begin{align}
\overline{G}^\text{(E)}_d[X-X'] = \int \frac{\dd^d k}{(2\pi)^d} \frac{e^{+i\vec{k} \cdot (\vec{X}-\vec{X}')}}{\vec{k}^2+m^2}
\end{align}
for $d=1,2$, followed by applying the recursion relations to obtain the results for higher dimensions.

Let us now choose $\langle X^d \vert m^2 \rangle = X^d$ and see that inserting it into the general formula eq. \eqref{GreensFunction_GeneralFormula} allows us to obtain $\overline{G}_d$ from $\overline{G}_{d+1}$. This example illustrates that regularity of the charge density $\langle y' \vert P_0^2 m^2 \rangle$ is not a necessary criteria in the general formula eq. \eqref{GreensFunction_GeneralFormula}. (We will check our assertion here for the Minkowski Green's functions; the Euclidean case follows the same steps.) Using the Fourier representation of the Green's function and denoting $k_\mu = (k_\perp, k_d)$,
\begin{align}
\int_{-\infty}^{\infty} \dd X'^d \frac{X'^d}{X^d} \overline{G}_{d+1}[X-X']
&= \int_{-\infty}^{\infty} \dd X'^d \frac{X'^d}{X^d} \int \frac{\dd^d k_\perp\dd k_d}{(2\pi)^{d+1}} \frac{e^{-ik_\perp\cdot(X_\perp-X'_\perp)} e^{ik_d(X^d-X'^d)}}{-k_\perp^2 + k_d^2 + m^2} .
\end{align}
By performing the $X'^d$-integral first, one would obtain a $2\pi i \delta'[k_d]$. Integrating-by-parts the $k_d$-derivative on the $\delta$-function,
\begin{align}
\int_{-\infty}^{\infty} \dd X'^d \frac{X'^d}{X^d} \overline{G}_{d+1}[X-X']
&= \frac{-i}{X^d} \int \frac{\dd^d k_\perp\dd k_d}{(2\pi)^d} \left( i X^d - \frac{2k_d}{-k_\perp^2+k_d^2+m^2} \right)\frac{e^{-ik_\perp\cdot(X_\perp-X'_\perp)}}{-k_\perp^2 + k_d^2 + m^2} e^{ik_d X^d} \delta[k_d] 
\end{align}
whose right hand side is the Fourier representation of $\overline{G}_d[X-X']$ once the $k_d$-integration is carried out.

\section{$(d \geq 2)$-Dimensional de Sitter Spacetime}
\label{Section_deSitter}

The hyperboloid lying outside the light cone of the origin $0^\mathfrak{A}$ in $(d+1)$ dimensional Minkowski spacetime, described by the relation
\begin{align}
-\eta_{\mathfrak{A}\mathfrak{B}} X^\mathfrak{A} X^\mathfrak{B} \equiv -X^2 = \frac{1}{H^2}, \qquad H > 0,
\end{align}
defines the $d$ dimensional de Sitter spacetime with Hubble parameter $H$. Let us employ hyperbolic/spherical coordinates to parametrize the region of Minkowski for $-X^2 \geq 0$: let $\tau\in\mathbb{R}$ be the time coordinate; $\rho \geq 0$ control the size of a given hyperboloid $(-X^2 = \rho^2)$; $\{\theta^i\}$ be the $(d-1)$ angular coordinates on a $(d-1)$-sphere; and $\widehat{n}[\vec{\theta}]$ be the unit radial (spatial) vector. Then
\begin{align}
\label{deSitter_X_parametrization}
X^\mathfrak{A}[\rho,\tau,\vec{\theta}] = \rho \left( \sinh[\tau], \cosh[\tau] \widehat{n}[\vec{\theta}] \right) .
\end{align}
The Minkowski geometry, for $-X^2 > 0$, is now
\begin{align}
\label{Embedding_deSitter}
\dd s^2 &= \rho^2 \left( \dd\tau^2 - \cosh^2[\tau] \dd\Omega_{d-1}^2 \right) - \dd\rho^2 ,
\end{align}
where $\dd\Omega_{d-1}^2$ is the $(d-1)$D sphere metric, such that the induced geometry on the $\rho=1/H$ surface is de Sitter spacetime:
\begin{align}
g_{\mu\nu}^\text{(dS)} \dd x^\mu \dd x^\nu
= \frac{1}{H^2} \left( \dd\tau^2 - \cosh^2[\tau] \dd\Omega_{d-1}^2 \right) .
\end{align}
These are known as the closed slicing coordinates, because constant time surfaces describe a closed sphere, and they cover the whole of de Sitter spacetime.

The de Sitter Green's functions will be expressed in terms of the O$[d,1]$ invariant object
\begin{align}
\label{deSitter_Z_Definitions}
Z[x,x'] \equiv \left.\Big( H^2 X\left[ \rho, x \right] \cdot X'\left[ \rho', x' \right] \Big)\right\vert_{\rho=\rho'=H^{-1}}.
\end{align}
With the closed slicing coordinates of eq. \eqref{deSitter_X_parametrization} we have
\begin{align}
Z\left[ \tau,\vec{\theta};\tau',\vec{\theta}' \right] = \sinh[\tau] \sinh[\tau'] - \cosh[\tau] \cosh[\tau'] \widehat{n}\cdot\widehat{n}' .
\end{align}
We recall that half the square of the geodesic distance between $x$ and $x'$ (Synge's world function) in de Sitter is 
\begin{align}
\label{deSitter_WorldFunction}
\sigma^\text{(dS)}[x,x'] = \frac{1}{2}\left(\frac{1}{H} \cosh^{-1}\Big[-Z[x,x']\Big] \right)^2.
\end{align}
As a consequence, the observer at $x$ and source at $x'$ lie precisely on or within each other's null cones, $\sigma^\text{(dS)}[x,x'] \geq 0$, and are thus causally connected, when and only when
\begin{align}
\label{deSitter_CausalCondition}
Z[x,x'] \leq -1 .
\end{align}

{\bf Integral representation} \qquad With the identification
\begin{align}
P[\rho] = \rho, \qquad g_\text{AB}^\perp \dd y^\text{A} \dd y^\text{B} = -\dd\rho^2 .
\end{align}
we may now invoke the general formula \eqref{GreensFunction_GeneralFormula} for the de Sitter embedding in eq. \eqref{Embedding_deSitter}. The eigenvector equation \eqref{EigenvectorEquation} we need to solve is
\begin{align}
-\frac{1}{\rho^{d-2}} \frac{\dd}{\dd \rho} \left(\rho^d \frac{\dd \langle \rho | (m/H)^2 \rangle}{\dd \rho}\right)
+ (\rho m)^2 \langle \rho | (m/H)^2 \rangle = \left(\frac{m}{H}\right)^2 \langle \rho | (m/H)^2 \rangle
\end{align}
and the two independent solutions are
\begin{align}
\frac{I_{\nu_\pm}[m \rho]}{(m \rho)^{\frac{d-1}{2}}}, \qquad 
\nu_\pm \equiv \pm\sqrt{\left(\frac{d-1}{2}\right)^2-\left(\frac{m}{H}\right)^2},
\end{align}
where $I_{\nu_\pm}$ is the modified Bessel function.\footnote{When $\nu_\pm$ is an integer, the Bessel $I_{\nu_\pm}$ are no longer independent. This will not concern us because, as already discussed previously, we really only need one solution of eq. \eqref{EigenvectorEquation}. The use of both $I_{\nu_\pm}$ when $\nu_\pm$ are not integers, is primarily a check of the calculation itself: both must lead to the same de Sitter Green's function.} For any complex $\nu$, $I_\nu[z]$ blows up as $z \to +\infty$ (see 8.451 of \cite{G&S}); and for negative real $\nu$, $I_\nu[z]$ blows up as $z \to 0^+$ (see 8.445 of \cite{G&S}). ($I_\nu[z]$ is also usually defined with a branch cut along the negative real line.) Therefore when $\nu_\pm$ is real, $(d-1)^2/4 \geq (m/H)^2$, it is impossible to find eigenvector solutions to eq. \eqref{EigenvectorEquation} that are regular everywhere. Similar remarks apply to the $d$-sphere problem in section \eqref{Section_dSphere} below.

The integral representation of the massive scalar Green's function in de Sitter spacetime is therefore
\begin{align}
\label{GreensFunction_deSitter_IntegralRepresentation}
G_d[x,x']
= \int_0^\infty \dd\rho' \left(H \rho'\right)^{\frac{d-3}{2}} \frac{I_{\nu_\pm}[m \rho']}{I_{\nu_\pm}[m/H]} \overline{G}_{d+1} \Big[ X[\rho=H^{-1},x] - X'[\rho',x'] \Big] ,
\end{align}
with the parametrization in eq. \eqref{deSitter_X_parametrization}, and the massive Green's functions $\overline{G}_{d+1}$ from equations \eqref{GreensFunction_Minkowski_RetardedAdvanced} through \eqref{GreensFunction_Minkowski_Odd}. Just as was for the massless case, to emulate the scalar field of a spacetime point source in the observer's de Sitter world, the higher dimensional experimentalists have laid down a line source which begins from one end very close to the Minkowski origin $0^\mathfrak{A}$, penetrating the de Sitter hyperboloid perpendicularly at $x'$, and extending to spatial infinity. (That the integral in eq. \eqref{GreensFunction_deSitter_IntegralRepresentation} does not include the origin, i.e., $\rho' =0$, will be clarified in the discussion after equations \eqref{ModifiedBesselEquation_Even} and \eqref{ModifiedBesselEquation_Odd}.)

\begin{figure}
\begin{center}
\includegraphics[width=3.2in]{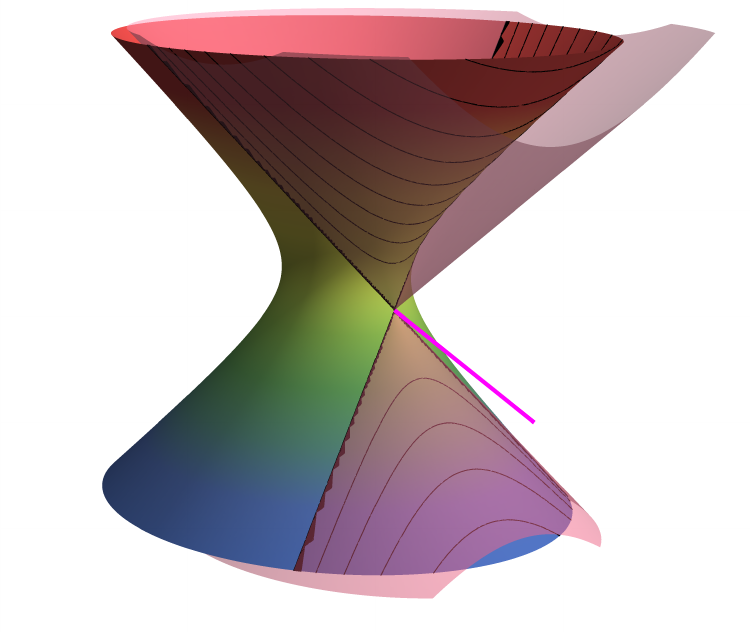}
\includegraphics[width=3.2in]{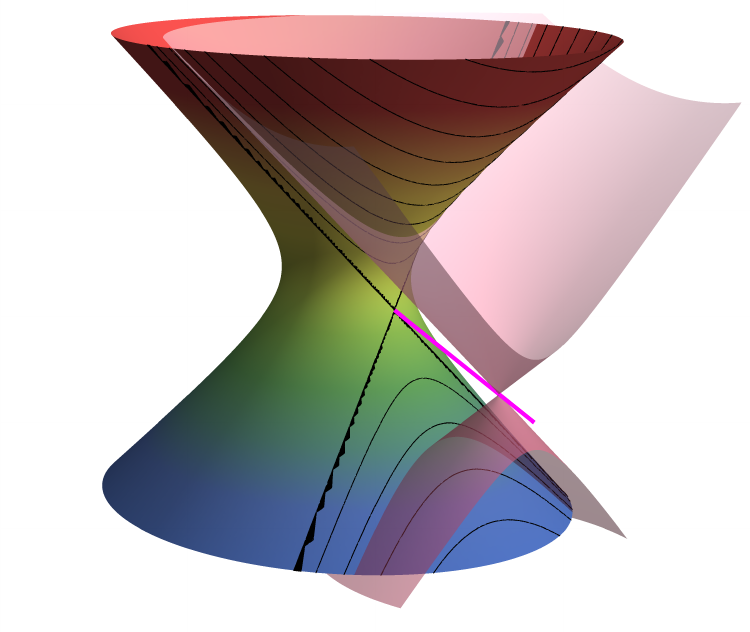}
\caption{On both panels, the hyperboloids represent de Sitter spacetime embedded in one higher dimensional Minkowski spacetime. The bold (slightly jagged) lines on the hyperboloids define the light cone of the point source of the massive scalar symmetric Green's function in de Sitter. (That is, the point source is located at the apex of the light cone.) The thin curved lines on the hyperboloids are the equipotential lines of the tail portion of the same Green's function. From the ambient Minkowski perspective, what sources the Green's function on the hyperboloid is the infinite line charge seen penetrating it at the location of the de Sitter point source. The translucent cones are ambient Minkowski light cones of the spacetime events located at their apexes. These light cones, when based at some location on the line charge, illustrate the relationship between the causal structure of signals generated in the ambient Minkowski spacetime and that of waves propagating on the de Sitter hyperboloid. {\it Left panel}: The intersection between the ambient light cone of the point source of the Green's function and that of the hyperboloid itself, is in fact the de Sitter light cone of the same point source. From the ambient perspective, the de Sitter point source lying on the line charge is entirely responsible for the light cone part of the Green's function; moreover, we have shown in \cite{Chu:2013hra} that it does not contribute to the tail part. {\it Right panel}: The intersection between the ambient light cone of some other point on the line charge and that of the hyperboloid, always lies within the de Sitter light cone of the Green's function's point source. From the ambient perspective, therefore, the tail part of the Green's function is completely due to signals from the rest of the line charge -- i.e., with its point of intersection with the hyperboloid removed.}
\label{Figure_CausalStructureFromEmbedding}
\end{center}
\end{figure}

{\bf Causal structure for closed slicing} \qquad We have shown in \cite{Chu:2013hra}, with the closed slicing coordinates of eq. \eqref{deSitter_X_parametrization}, that the retarded (advanced) Green's function in de Sitter Green's function can be obtained by utilizing the corresponding the retarded (advanced) Minkowski Green's function in eq. \eqref{GreensFunction_deSitter_IntegralRepresentation}:
\begin{align}
\label{GreensFunction_deSitter_CausalityOverlapsWithMinkowskiInClosedSlicing}
G_d^{(\text{Closed}\vert\pm)}[x,x'] 
&= \int_0^\infty \dd\rho' \left(H \rho'\right)^{\frac{d-3}{2}} \frac{I_{\nu_\pm}[m \rho']}{I_{\nu_\pm}[m/H]} 
		\Theta\left[ \pm(X^0-X'^0) \right] \bar{\mathcal{G}}_{d+1} \Big[ X[\rho,\tau,\widehat{n}] - X'[\rho',\tau',\widehat{n}'] \Big] \\
&\equiv \Theta[\pm(\tau-\tau')] \mathcal{G}_d[x,x'] ,
\end{align}
where $\bar{\mathcal{G}}_{d+1}$ and $\mathcal{G}_d$ are both symmetric (i.e., un-ordered in time); the former can be found in equations \eqref{GreensFunction_Minkowski_Even} and \eqref{GreensFunction_Minkowski_Odd}, and the latter is:
\begin{align}
\label{GreensFunction_deSitter_SymmetricDefinition}
\mathcal{G}_d[x,x'] &\equiv \int_0^\infty \dd\rho' \left(H \rho'\right)^{\frac{d-3}{2}} \frac{I_{\nu_\pm}[m \rho']}{I_{\nu_\pm}[m/H]} 
 \bar{\mathcal{G}}_{d+1} \Big[ X[\rho=H^{-1},\tau,\widehat{n}] - X'[\rho',\tau',\widehat{n}'] \Big] .
\end{align}
That the retarded (advanced) Green's function in the ambient Minkowski generates the retarded (advanced) counterpart in de Sitter can be seen by exploiting the SO$[d,1]$ symmetry of the setup \cite{Chu:2013hra}. (The argument is independent of the mass $m$.) If $X$ lies on the de Sitter hyperboloid and $X'$ lies on the line source parametrized in eq. \eqref{deSitter_X_parametrization}, then as far as the study of their chronology is concerned, we may perform a global Lorentz boost so that $X' = \rho'(0,\widehat{n}'[\vec{\theta}'])$, i.e., $X'$ is now purely spatial, with the associated de Sitter time $\tau'=0$. The retarded (advanced) signal produced by such an $X'$ can thus only reach the positive (negative) half of the de Sitter hyperboloid, $X^0 \geq 0$ ($X^0 \leq 0$), which then translates to $\tau \geq \tau'$ ($\tau \leq \tau'$). The analysis in \cite{Chu:2013hra} also revealed that, for a fixed observer location $X[\rho=1/H,x]$ on the de Sitter hyperboloid, the only point on the line source in eq. \eqref{deSitter_X_parametrization} satisfying the de Sitter light cone condition $\sigma^\text{(dS)}[x,x'] = 0 \Leftrightarrow Z[x,x'] = -1$ is the one where $\rho' = 1/H$; this allowed us to identify, from the ambient Minkowski perspective, that the source of the light cone part of the de Sitter Green's function to be the location where the line source in eq. \eqref{deSitter_X_parametrization} intersects the de Sitter hyperboloid at $X'[\rho'=1/H,x']$. The tail portion of $G_d[x,x']$ is, in turn, sourced by the rest of the line satisfying $0 < H^2 \sbar = -(1/2)(H^2 \rho'^2 + 1 + 2 H \rho' Z)$. For a visual representation of this causal structure, see Fig.  \eqref{Figure_CausalStructureFromEmbedding}; also consult Fig. (1) of \cite{Chu:2013hra} for a plot of the relevant portion of the line source in eq. \eqref{deSitter_X_parametrization} contributing to the scalar signal at $X$.

In what follows, we will focus on computing the symmetric Green's function in eq. \eqref{GreensFunction_deSitter_SymmetricDefinition}; the retarded (or advanced) Green's function written in any coordinate system can then be obtained by multiplying $\mathcal{G}_d[x,x']$ by $\Theta[t-t']$ (or $\Theta[t'-t]$), where $t$ and $t'$ are the relevant time coordinates.

{\bf Technicalities} \qquad We will now set $H=1$ for convenience. In eq. \eqref{GreensFunction_deSitter_SymmetricDefinition}, Synge's world function now takes the expression
\begin{align}
\sbar = -\frac{1}{2} \left( \rho'^2 + 1 + 2 \rho' Z \right)
\end{align}
Recalling the presence of $\Theta[\sbar]$ in equations \eqref{GreensFunction_Minkowski_Even} and \eqref{GreensFunction_Minkowski_Odd}, ensuring no signals travel outside the Minkowski light cone, we see that the limits of $\rho'$ integration are restricted to the range $\rho' \in[-Z-\sqrt{Z^2-1}, -Z+\sqrt{Z^2-1}]$. If we further perform a change-of-variables from differentiating with respect to $\sbar$, to that with respect to $Z$, then we shall find that
\begin{align}
\label{GreensFunction_deSitter_Even_Integral}
\mathcal{G}_{\text{even }d}[x,x']
&= \frac{1}{(2\pi)^{\frac{d}{2}} I_{\nu_\pm}[m]} \left( -\frac{\partial}{\partial Z} \right)^{\frac{d-2}{2}} 
\left(\Theta[-Z-1] \int_{-Z-\sqrt{Z^2-1}}^{-Z+\sqrt{Z^2-1}} \frac{\dd\rho'}{\sqrt{\rho'}} \frac{I_{\nu_\pm}[m \rho'] \cos\left[ m \sqrt{2 \bar{\sigma}} \right]}{\sqrt{2 \bar{\sigma}}} \right)  \\
\label{GreensFunction_deSitter_Odd_Integral}
\mathcal{G}_{\text{odd }d}[x,x']
&= \frac{1}{2(2\pi)^{\frac{d-1}{2}} I_{\nu_\pm}[m]} \left( -\frac{\partial}{\partial Z} \right)^{\frac{d-1}{2}}
\left(\Theta[-Z-1] \int_{-Z-\sqrt{Z^2-1}}^{-Z+\sqrt{Z^2-1}} \frac{\dd\rho'}{\rho'} I_{\nu_\pm}[m \rho'] J_0\left[m\sqrt{2\bar{\sigma}}\right] \right) .
\end{align}
To evaluate these integrals, we remember the expectation that $I_{\nu_\pm}[m]$ needs to cancel out of the final answer. One may verify this by first noting that the modified Bessel function obeys the ordinary differential equation
\begin{align}
\label{ModifiedBesselEquation}
\mathfrak{D}_z^{(\nu)} I_\nu[z] \equiv z^2 I''_\nu[z] + z I'_\nu[z] - (z^2 + \nu^2) I_\nu[z] = 0 .
\end{align}
If we replace $\nu_\pm \to \nu$ and view $\nu$, for the moment, as independent of $d$ and $m$, and if we take into account eq. \eqref{ModifiedBesselEquation} obeyed by $I_\nu$, then one may deduce via a direct calculation that
\begin{align}
\label{ModifiedBesselEquation_Even}
\mathfrak{D}_m^{(\nu)} \left(\int_{-Z-\sqrt{Z^2-1}}^{-Z+\sqrt{Z^2-1}} \frac{\dd\rho'}{\sqrt{\rho'}} \frac{I_\nu [m \rho'] \cos\left[ m \sqrt{2 \bar{\sigma}} \right]}{\sqrt{2 \bar{\sigma}}}\right)
&= -2m \int_{-Z-\sqrt{Z^2-1}}^{-Z+\sqrt{Z^2-1}} \dd\rho' \frac{\partial}{\partial \rho'}
\left( \sqrt{\rho'} \sin\left[m\sqrt{2\bar{\sigma}}\right] I_\nu[m\rho'] \right),
\end{align}
and
\begin{align}
\label{ModifiedBesselEquation_Odd}
\mathfrak{D}_m^{(\nu)} \left(\int_{-Z-\sqrt{Z^2-1}}^{-Z+\sqrt{Z^2-1}} \frac{\dd\rho'}{\rho'} I_\nu[m \rho'] J_0\left[m\sqrt{2\bar{\sigma}}\right]\right)
&= 2m \int_{-Z-\sqrt{Z^2-1}}^{-Z+\sqrt{Z^2-1}} \dd\rho' \frac{\partial}{\partial \rho'}
\left( \sqrt{2\bar{\sigma}} I_\nu[m\rho'] J'_0\left[m\sqrt{2\bar{\sigma}}\right] \right) .
\end{align}
Because the limits of integration are precisely the zeros of $\bar{\sigma}$, the integrals on the right hand sides of equations \eqref{ModifiedBesselEquation_Even} and \eqref{ModifiedBesselEquation_Odd} are zero -- observe that the lower limit of the $\rho'$ integral is never zero, for $Z \leq -1$, unless $Z = -\infty$. We therefore do not need to worry about evaluating $I_\nu[0]$, which may otherwise be singular for $\nu \leq 0$. This immediately implies the integrals in equations \eqref{GreensFunction_deSitter_Even_Integral} and \eqref{GreensFunction_deSitter_Odd_Integral} are a linear combination of $I_{\pm\nu}[m]$, with coefficients that do not depend on $m$. On the other hand, because the range of integration in equations \eqref{GreensFunction_deSitter_Even_Integral} and \eqref{GreensFunction_deSitter_Odd_Integral} is finite, we may Taylor expand their integrands as a power series in $m$. Now, 8.445 and 8.441.1 of \cite{G&S} inform us that $I_\nu[z]$ is $z^\nu$ multiplied by an even power series in $z$ whereas $J_0[z]$ (and $\cos[z]$) is purely an even power series in $z$. This indicates the integrals in equations \eqref{GreensFunction_deSitter_Even_Integral} and \eqref{GreensFunction_deSitter_Odd_Integral} are themselves $m^\nu$ times an even power series in $m$, which therefore has to be proportional to $I_\nu[m]$ itself, since the other solution, $I_{-\nu}[m]$, is $m^{-\nu}$ times an even power series in $m$. (This argument breaks down for discrete values of $\nu$, but can be saved by assuming continuity in the parameter $\nu$.) 

To sum, we have managed to argue that the integrals in equations \eqref{GreensFunction_deSitter_Even_Integral} and \eqref{GreensFunction_deSitter_Odd_Integral} are both $I_\nu[m]$ multiplied by a $m$-independent coefficient. Since the coefficient does not depend on $m$, we may now evaluate it by setting $m=0$ while holding $\nu$ fixed in equations \eqref{ModifiedBesselEquation_Even} and \eqref{ModifiedBesselEquation_Odd}, namely
\begin{align}
\label{GreensFunction_deSitter_Even_Integral_II}
\int_{-Z-\sqrt{Z^2-1}}^{-Z+\sqrt{Z^2-1}} \frac{\dd\rho'}{\sqrt{\rho'}} \frac{I_\nu[m \rho'] \cos\left[ m \sqrt{2 \bar{\sigma}} \right]}{I_\nu[m] \sqrt{2 \bar{\sigma}}}
&= \int_{-Z-\sqrt{Z^2-1}}^{-Z+\sqrt{Z^2-1}} \frac{\dd\rho'}{\sqrt{\rho'}} \lim_{\substack{m \to 0 \\ \nu \text{ fixed}}} \frac{I_\nu[m \rho'] \cos\left[ m \sqrt{2 \bar{\sigma}} \right]}{I_\nu[m] \sqrt{2 \bar{\sigma}}} \\
\label{GreensFunction_deSitter_Odd_Integral_II}
\int_{-Z-\sqrt{Z^2-1}}^{-Z+\sqrt{Z^2-1}} \frac{\dd\rho'}{\rho'} \frac{I_\nu[m \rho']}{I_\nu[m]} J_0\left[m\sqrt{2\bar{\sigma}}\right]
&= \int_{-Z-\sqrt{Z^2-1}}^{-Z+\sqrt{Z^2-1}} \frac{\dd\rho'}{\rho'} \lim_{\substack{m \to 0 \\ \nu \text{ fixed}}} \frac{I_\nu[m \rho']}{I_\nu[m]} J_0\left[m\sqrt{2\bar{\sigma}}\right] .
\end{align}
Taking these limits requires knowing that $J_0[0] = 1$ and, for $|z| \ll 1$,
\begin{align}
I_\nu[z] = \frac{(z/2)^\nu}{\Gamma[\nu+1]}\left(1 + \mathcal{O}[z^2] \right) .
\end{align}
($\Gamma$ is the Gamma function.) Putting back the $\nu_\pm$, we surmise at this point, that
\begin{align}
\label{GreensFunction_deSitter_Even_Integral_III}
\mathcal{G}_{\text{even }d}[x,x']
&= \frac{1}{(2\pi)^{\frac{d}{2}}} \left( -\frac{\partial}{\partial Z} \right)^{\frac{d-2}{2}} 
\left(\Theta[-Z-1] \int_{-Z-\sqrt{Z^2-1}}^{-Z+\sqrt{Z^2-1}} \frac{\dd\rho' \ \rho'^{\nu_\pm-\frac{1}{2}}}{\sqrt{2 \bar{\sigma}}} \right) \\
\label{GreensFunction_deSitter_Odd_Integral_III}
\mathcal{G}_{\text{odd }d}[x,x']
&= \frac{1}{2(2\pi)^{\frac{d-1}{2}}} \left( -\frac{\partial}{\partial Z} \right)^{\frac{d-1}{2}}
\left(\Theta[-Z-1] \int_{-Z-\sqrt{Z^2-1}}^{-Z+\sqrt{Z^2-1}} \dd\rho' \rho'^{\nu_\pm-1} \right) .
\end{align}
The integral in eq. \eqref{GreensFunction_deSitter_Even_Integral_III} may be handled by changing variables to $\rho' \equiv -Z + \cos[u] \sqrt{Z^2-1}$ (with $u \in [0,\pi]$), followed by recognizing the integral representation of the Legendre function (see 8.822.1 of \cite{G&S}). For odd $d$, we will set $\delta[-Z-1] \sinh\left[\nu\ln\left[-Z+\sqrt{Z^2-1}\right]\right] \to 0$, so that the result after evaluating eq. \eqref{GreensFunction_deSitter_Odd_Integral_III} has only $(d-3)/2$ (and not $(d-1)/2$) derivatives acting on it.

Comparing equations \eqref{GreensFunction_deSitter_Even_Integral_III} and \eqref{GreensFunction_deSitter_Odd_Integral_III} with equations (76) and (77) of \cite{Chu:2013hra} tells us that the only difference between the two pairs of equations is the exponent of $\rho'$ within the integrands. In particular, this means we could have arrived at equations \eqref{GreensFunction_deSitter_Even_Integral_III} and \eqref{GreensFunction_deSitter_Odd_Integral_III} using the massless Minkowski Green's function in equations (37)-(39) of \cite{Chu:2013hra}, but replacing $(H \rho')^{d-2}$ in its eq. (61) with $(H \rho')^{\nu_\pm + \frac{d-3}{2}}$. Physically, this means the higher dimensional Minkowski beings may set up appropriate charge densities that produce massless scalar waves to deceive the de Sitter observer she is detecting massive scalar waves due to a spacetime point charge in her own world.

{\bf Results} \qquad We have arrived at the following results for the retarded $G^+_d[x,x']$ and advanced $G^-_d[x,x']$ Green's functions of $\Box + m^2$ in $(d \geq 2)$-dimensional de Sitter spacetime. In terms of the closed slicing coordinates found in eq. \eqref{deSitter_X_parametrization},
\begin{align}
\label{GreensFunction_deSitter_Closed}
G_d^{(\text{Closed}\vert\pm)}[x,x'] = \Theta[\pm(\tau-\tau')] \mathcal{G}_d[x,x'] .
\end{align}
With $Z[x,x']$ defined in eq. \eqref{deSitter_Z_Definitions}, and restoring $H$, the symmetric Green's functions $\mathcal{G}_d[x,x']$ are
\begin{align}
\label{GreensFunction_deSitter_Even}
\mathcal{G}_{\text{even }d}[x,x']
&= \frac{\pi H^{d-2}}{(2\pi)^{\frac{d}{2}}} \left( -\frac{\partial}{\partial Z} \right)^{\frac{d-2}{2}} 
\left(\Theta[-Z-1] P_{\nu-\frac{1}{2}}\left[-Z\right] \right) \\
\label{GreensFunction_deSitter_Odd}
\mathcal{G}_{\text{odd }d}[x,x']
&= \frac{H^{d-2}}{(2\pi)^{\frac{d-1}{2}}} \left( -\frac{\partial}{\partial Z} \right)^{\frac{d-3}{2}}
\left( \frac{\Theta[-Z-1]}{\sqrt{Z^2-1}} \cosh\left[\nu\ln\left[-Z+\sqrt{Z^2-1}\right]\right] \right) ,
\end{align}
($P_{\nu-1/2}$ is the Legendre function) and
\begin{align}
\label{nu}
\nu \equiv \sqrt{\left(\frac{d-1}{2}\right)^2-\left(\frac{m}{H}\right)^2} .
\end{align}
The tail portion of these Green's functions can be read off directly:
\begin{align}
\label{GreensFunction_deSitter_Even_Tail}
\mathcal{G}^\text{(Tail)}_{\text{even }d}[x,x']
&= \frac{\pi H^{d-2}}{(2\pi)^{\frac{d}{2}}} \Theta[-Z-1] \left( -\frac{\partial}{\partial Z} \right)^{\frac{d-2}{2}} P_{\nu-\frac{1}{2}}\left[-Z\right] \\
\label{GreensFunction_deSitter_Odd_Tail}
\mathcal{G}^\text{(Tail)}_{\text{odd }d}[x,x']
&= \frac{H^{d-2}}{(2\pi)^{\frac{d-1}{2}}} \Theta[-Z-1] \left( -\frac{\partial}{\partial Z} \right)^{\frac{d-3}{2}}
	\frac{\cosh\left[\nu\ln\left[-Z+\sqrt{Z^2-1}\right]\right]}{\sqrt{Z^2-1}}
\end{align}
Notice the results in equations \eqref{GreensFunction_deSitter_Even} and \eqref{GreensFunction_deSitter_Odd} are invariant under the replacement $\nu \to -\nu$; to see this for even $d$, one needs the identity $P_\mu[z] = P_{-\mu-1}[z]$. This confirms the expectation that the two charge densities associated with $I_{\nu_\pm}[m\rho']/(m\rho')^{\frac{d-1}{2}}$ give the same final result for the Green's function. Moreover, this also informs us that, even when $\nu$ becomes purely imaginary -- when the mass becomes large enough such that $(m/H)^2 > (d-1)^2/4$ -- the Green's function is still a purely real object.\footnote{A check of our results here is to ensure that the massive wave operator in de Sitter spacetime annihilates the tail part of the Green's functions, the expressions multiplying $\Theta[-Z-1]$ in equations \eqref{GreensFunction_deSitter_Even_Tail} and \eqref{GreensFunction_deSitter_Odd_Tail}. We have done so for $d=2,3,\dots,13$ on the computer \cite{Mathematica}. Because the Green's function depends on $x$ and $x'$ solely through the object $Z$, the wave operator is built out of $Z$-derivatives, $(\Box_{(d)} + m^2)G = -(1-Z^2)G''[Z] + d Z G'[Z] + m^2 G[Z]$ -- see for e.g., eq. (37) of \cite{Spradlin:2001pw}; the $P$ there is our $-Z$ here, and they are using the `mostly plus' convention for $\eta_{\mathfrak{A}\mathfrak{B}}$. Further note that, the only occurrence of mass in the solutions of equations \eqref{GreensFunction_deSitter_Even} and \eqref{GreensFunction_deSitter_Odd} is in $\nu$ defined in eq. \eqref{nu}. Together with the presence of multiple $-Z$ derivatives in equations \eqref{GreensFunction_deSitter_Even} and \eqref{GreensFunction_deSitter_Odd}, this indicates starting from the $d=2,3$ solutions and up to an overall multiplicative constant, the $d+2$ dimensional solution $G_{d+2}$ can be gotten from the $d$ dimensional one $G_d$ by first shifting $m^2 \to m^2 - d$ (so that $\nu = ((d-1)^2/4 - m^2)^{1/2} \to ((d+1)^2/4 - m^2)^{1/2}$) and then acting on the latter with $\partial/\partial(-Z)$. This can be understood by a direct computation using the form of $\Box$ in terms of $Z$; specifically, for $x \neq x'$, one finds $0 = (\partial/\partial Z)(\Box_{(d)} + m^2 - d) G_d[m^2 \to m^2-d] = (\Box_{d+2} + m^2)(\partial/\partial Z) G_d[m^2 \to m^2-d]$. We have also checked that the $d=4$ result coincides with eq. D1 of \cite{Rosenthal:2003qr} after converting the $\delta[\sigma]$ there to $\delta[Z+1]/|\partial \sigma[Z=-1]/Z|$; in \cite{Rosenthal:2003qr}, set $\lambda=1$, $\sigma = \sigma^{\text{(dS)}}$ (see eq. \eqref{deSitter_WorldFunction}) and $s=\sqrt{2\sigma^{\text{(dS)}}}$, for ease of comparison. (In this footnote, we have set $H=1$ for convenience.)}

{\bf Conformal Flatness} \qquad For arbitrary $d \geq 2$, de Sitter spacetime is conformally flat, while the scalar action
\begin{align}
\label{ConformallyInvariantAction}
S_\varphi &= \int \dd^d x \frac{\sqrt{|g|}}{2} \left( \left(\nabla\varphi\right)^2 + \frac{d-2}{4(d-1)} \mathcal{R}[g] \varphi^2 \right)
\end{align}
is invariant (up to a surface term) if we simultaneously replace
\begin{align}
g_{\mu\nu} \to \Omega^2 g_{\mu\nu}, \qquad \varphi \to \Omega^{1-\frac{d}{2}} \varphi .
\end{align}
In de Sitter spacetime the Ricci scalar is 
\begin{align}
\mathcal{R} = -d(d-1)H^2,
\end{align}
so that the ``mass term" becomes $\frac{d-2}{4(d-1)} \mathcal{R}[g] \varphi^2 = d(d-2)(H/2)^2 \varphi^2$. This means if we choose (mass)$^2$ to be exactly
\begin{align}
m^2 \to \frac{d(d-2)}{4} H^2,
\end{align}
thereby setting $\nu = 1/2$ in equations \eqref{GreensFunction_deSitter_Even} and \eqref{GreensFunction_deSitter_Odd}, our scalar Green's functions should transform into the massless Green's functions in $d$-dimensional Minkowski, multiplied by $(\Omega[x]\Omega[x'])^{1-(d/2)}$. The conformal factor $\Omega$ can, in turn, be worked out using the embedding defined by:
\begin{align*}
X^0 = \frac{1}{2\eta} \left( \eta^2 - \vec{x}^2 - \frac{1}{H^2} \right), \qquad
X^d = \frac{1}{2\eta} \left( -\eta^2 + \vec{x}^2 - \frac{1}{H^2} \right), \\
X^i = \frac{x^i}{H \eta}, \qquad i = 1,2,\dots,d-1, \qquad \eta < 0.
\end{align*}
The de Sitter geometry now takes the expression
\begin{align*}
g_{\mu\nu}^\text{(dS)} \dd x^\mu \dd x^\nu = \Omega^2[\eta] \left(\dd\eta^2 - \dd\vec{x}^2\right), \qquad
\Omega[\eta] = \frac{1}{H\eta} .
\end{align*}
Let us abuse notation somewhat and define
\begin{align}
\sbar \equiv \frac{1}{2} \left( (\eta-\eta')^2 - (\vec{x}-\vec{x}')^2 \right) ,
\end{align}
in terms of which
\begin{align}
-Z-1 = \frac{\sbar}{\eta\eta'} .
\end{align}
For even $d$, $P_{\nu-1/2}[-Z] \to P_0[-Z] = 1$. Next, if we use the property that $\eta\eta' > 0$ to re-write $\Theta[-Z-1] = \Theta[\sbar]$, and then proceed to differentiate with respect to $(\eta\eta')(-Z-1) = \sbar$ instead of $-Z$,
\begin{align}
\label{GreensFunction_deSitter_Even_ConformalLimit}
\mathcal{G}_{\text{even }d}[x,x']
&= \frac{1}{\Omega[\eta]^{\frac{d}{2}-1} \Omega[\eta']^{\frac{d}{2}-1}} \cdot 
\frac{1}{2(2\pi)^{\frac{d}{2}-1}} \left( \frac{\partial}{\partial \bar{\sigma}} \right)^{\frac{d-2}{2}} \Theta[\bar{\sigma}] 
\end{align}
For odd $d$, it is useful to re-define $-Z \equiv \cosh[2\chi]$ for $\chi \geq 0$. (Remember $-Z \geq 1$.) This means $\cosh\left[\nu\ln\left[-Z+\sqrt{Z^2-1}\right]\right] = \sqrt{(1+\cosh[2\chi])/2} = \sqrt{(1-Z)/2}$. Manipulations similar to the even $d$ case then hands us
\begin{align}
\label{GreensFunction_deSitter_Odd_ConformalLimit}
\mathcal{G}_{\text{odd }d}[x,x']
&= \frac{1}{\Omega[\eta]^{\frac{d}{2}-1} \Omega[\eta']^{\frac{d}{2}-1}} \cdot 
\frac{1}{\sqrt{2}(2\pi)^{\frac{d-1}{2}}} \left( \frac{\partial}{\partial \bar{\sigma}} \right)^{\frac{d-3}{2}}
\left( \frac{\Theta[\bar{\sigma}]}{\sqrt{\bar{\sigma}}} \right) 
\end{align}
Comparing equations \eqref{GreensFunction_deSitter_Even_ConformalLimit} and \eqref{GreensFunction_deSitter_Odd_ConformalLimit} with equations (38) and (39) of \cite{Chu:2013hra} indicates we have indeed verified that
\begin{align}
G^\text{(de Sitter)}_d\left[ x,x'; m^2 = d(d-2) (H/2)^2 \right]
= \left(\Omega[\eta] \Omega[\eta']\right)^{1-\frac{d}{2}} \overline{G}_d[\eta-\eta',\vec{x}-\vec{x}';m=0] .
\end{align}
where the $\overline{G}_d$ on the right hand side is the $m=0$ limit of equations \eqref{GreensFunction_Minkowski_RetardedAdvanced} through \eqref{GreensFunction_Minkowski_Odd}, with $X-X'$ replaced with $(\eta-\eta',\vec{x}-\vec{x}')$. That we were able to recover the conformal limit is a check on our Green's function results in equations \eqref{GreensFunction_deSitter_Even} and \eqref{GreensFunction_deSitter_Odd}.

\section{$(d \geq 2)$-Sphere}
\label{Section_dSphere}

In \cite{Chu:2013hra} we used the analog of the general formula in eq. \eqref{GreensFunction_GeneralFormula} to attempt a derivation of the Green's function of the Laplacian on the $d$-sphere. The motivation was to understand how the general formula would break down because we knew that no solution should exist: the massless scalar charge density $J \equiv \Box \varphi$ on a sphere should integrate to zero. (While the massless field generated by a single point source does not exist, the field generated by a pair of charges -- one positive and the other negative -- does exist, and we used the general formula to guide us to the answer.) In this section, we shall show that the massless limit is a discontinuous one, because it is equivalent to setting the radius $R$ of the $d$-sphere to zero,\footnote{$R$ is not to be confused with the $\bar{R} \equiv |\vec{X}-\vec{X}'|$ in equations \eqref{GreensFunction_Euclidean_Even} and \eqref{GreensFunction_Euclidean_Odd}.} i.e., it is the limit $R \ll m^{-1}$. Since topology no longer imposes any restriction on the integral of $J \equiv (-\Box + m^2)\varphi$ on the $d$-sphere, we will in fact use the general formula eq. \eqref{GreensFunction_GeneralFormula} to derive the Green's function of the Helmholtz operator $-\Box + m^2$.

The Euclidean metric in $(d+1)$ dimensions written in spherical coordinates is
\begin{align}
-\left(\dd \vec{x}^2\right)_{d+1} = - \dd r^2 - r^2 \left( \dd\theta^2 + \sin^2[\theta] \dd\Omega_{d-1}^2 \right).
\end{align}
(We have multiplied both sides by a negative sign to account for the $-$ sign in front of the Laplacian, for ease of identification with the formalism we laid out in section \eqref{Section_Generalities}.) The situation of having a $d$-sphere of radius $R$ is analogous to that of the $d$-dimensional de Sitter case, except we replace $\rho \to r$ and identify inverse Hubble to be the radius of the $d$-sphere, $1/H \to R$. 

Let $\vec{\theta}$ be the $d$ angular coordinates denoting the position of the observer on the $d$ sphere and let $\vec{\theta}'$ be the position of the source of her Green's function. We will denote as $\widehat{n}$ and $\widehat{n}'$ the unit radial vectors associated with, respectively, the observer and source in the ambient Euclidean space. 

First, we need to solve the eigenvector equation 
\begin{align}
-\frac{1}{r^{d-2}} \frac{\dd}{\dd r}\left(r^d \frac{\dd \langle r \vert (mR)^2 \rangle}{\dd r} \right) + \left(mr\right)^2 \langle r \vert (mR)^2 \rangle = \left(mR\right)^2 \langle r \vert (mR)^2 \rangle .
\end{align}
The two independent charge densities are described by the solutions
\begin{align}
\frac{I_{\nu_\pm}[mr]}{(mr)^{\frac{d-1}{2}}}, \qquad 
\nu_\pm \equiv \sqrt{\left(\frac{d-1}{2}\right)^2-(mR)^2} .
\end{align}
and the corresponding line source in the $(d+1)$ dimensional Euclidean space is, for fixed $\vec{\theta}'$,
\begin{align}
\label{GreensFunction_dSphere_LineSource}
\vec{X}' = r'\widehat{n}[\vec{\theta}'] .
\end{align}
From equations \eqref{GreensFunction_GeneralFormula}, \eqref{GreensFunction_Euclidean_Even} and \eqref{GreensFunction_Euclidean_Odd}, the integral representations for the Green's function of the Helmholtz operator is
\begin{align}
\label{GreensFunction_dSphere_Even_Integral}
G_{\text{even } d}[\widehat{n}\cdot\widehat{n}';R]
&= \frac{1}{2m} \int_0^\infty \dd r' \left(\frac{r'}{R}\right)^{\frac{d-3}{2}} \frac{I_{\nu_\pm}[mr']}{I_{\nu_\pm}[mR]}
\left(-\frac{1}{2\pi \bar{R}} \frac{\partial}{\partial \bar{R}}\right)^{\frac{d}{2}} e^{-m \bar{R}}, \\
\label{GreensFunction_dSphere_Odd_Integral}
G_{\text{odd } d}[\widehat{n}\cdot\widehat{n}';R]
&= \frac{1}{2\pi} \int_0^\infty \dd r' \left(\frac{r'}{R}\right)^{\frac{d-3}{2}} \frac{I_{\nu_\pm}[mr']}{I_{\nu_\pm}[mR]}
\left(-\frac{1}{2\pi \bar{R}} \frac{\partial}{\partial \bar{R}}\right)^{\frac{d-1}{2}} K_0[m \bar{R}],
\end{align}
where
\begin{align}
\bar{R} = \sqrt{r'^2 + R^2 - 2 r' R \widehat{n}\cdot\widehat{n}'} .
\end{align}
For technical convenience, we will now set $R=1$. Converting the derivative with respect to $\bar{R}$ to one with respect to $\widehat{n}\cdot\widehat{n}'$,
\begin{align}
G_{\text{even } d}[\widehat{n}\cdot\widehat{n}']
&= \frac{1}{2 (2\pi)^{\frac{d}{2}}} 
	\left( \frac{\partial}{\partial \left(\widehat{n}\cdot\widehat{n}'\right)} \right)^{\frac{d}{2}} 
	\int_0^\infty  \frac{\dd r'}{r'^{\frac{3}{2}}} \frac{I_{\nu_\pm}[mr']}{I_{\nu_\pm}[m]} \frac{e^{-m \bar{R}}}{m}, \\
G_{\text{odd } d}[\widehat{n}\cdot\widehat{n}']
&= \frac{1}{(2\pi)^{\frac{d+1}{2}}} 
	\left(\frac{\partial}{\partial \left(\widehat{n}\cdot\widehat{n}'\right)}\right)^{\frac{d-1}{2}} 
	\int_0^\infty \frac{\dd r'}{r'} \frac{I_{\nu_\pm}[mr']}{I_{\nu_\pm}[m]} K_0[m \bar{R}] .
\end{align}
Recalling the discussion in section \eqref{Section_Generalities}, we expect the $I_{\nu_\pm}[m]$ to cancel out in the final result. We may attempt to verify this in the same manner as we did for the de Sitter case, i.e., replacing $\nu_\pm \to \nu$, regarding $\nu$ to be independent of $d$, $m$, $R$, and followed by applying (see eq. \eqref{ModifiedBesselEquation}), for odd $d$,
\begin{align*}
\mathfrak{D}^{(\nu)}_m \int_0^\infty \frac{\dd r'}{r'} I_\nu[mr'] K_0[m \bar{R}] 
&= -2m \int_0^\infty \dd r' \partial_{r'} \left( \bar{R} I_\nu[mr'] K_1[m \bar{R}] \right) 
\end{align*}
Using the large and small argument limits of the modified Bessel functions, which can be found in 8.445 and 8.451 of \cite{G&S} -- and if we further assume that $m > 0$ and Re$[\nu] > 0$,
\begin{align*}
\mathfrak{D}^{(\nu)}_m \int_0^\infty \frac{\dd r'}{r'} I_\nu[mr'] K_0[m \bar{R}] = -1
\end{align*}
Since, for $d \geq 3$, there is always at least one $\partial/\partial(\widehat{n}\cdot\widehat{n}')$ acting on this expression, the final answer is therefore indeed annihilated by $\mathfrak{D}^{(\nu)}_m$. However, since the integration range is now infinite, it is not possible to Taylor expand the integrand, integrate term-by-term, and argue that the resulting power series is proportional to $I_{\nu_\pm}[m]$.

Now for even $d$, if we assume $\text{Re}[\nu] > -1/2$ and $m>0$,
\begin{align}
\label{GreensFunction_dSphere_Even_Integral_CheckFail}
\mathfrak{D}^{(\nu)}_m \int_0^\infty \frac{\dd r'}{r'^{\frac{3}{2}}} I_\nu[mr'] \frac{\partial}{\partial(\widehat{n}\cdot\widehat{n}')} \frac{e^{-m \bar{R}}}{m}
= -2m \int_0^\infty \dd r' \partial_{r'} \left( \sqrt{r'} I_\nu[mr'] e^{-m \bar{R}} \right) 
= -\sqrt{\frac{2m}{\pi}} .
\end{align}
Notice, for $d=2$, we are unable to show that the integral of interest is annihilated by $\mathfrak{D}^{(\nu)}_m$, because there are no further derivatives with respect to $\widehat{n} \cdot \widehat{n}'$ acting on eq. \eqref{GreensFunction_dSphere_Even_Integral_CheckFail}.

We have just witnessed that these intermediate steps for the $d$-sphere calculation are significantly more subtle than the corresponding ones in de Sitter. In section \eqref{Section_deSitter}, the integrals converged without the need to make additional assumptions about $m$ or $\nu$. We were also able to successfully argue in some detail that the $I_\nu[m]$ did in fact cancel out of the final result. Here, we will resort to the more pragmatic route: we will assume that our general analysis in section \eqref{Section_Generalities} is correct, and assume that $I_\nu[m]$ does cancel for the $d$-sphere calculation at hand. At the end of the calculation, we will validate the final result by exercising various checks. 

Like the de Sitter case, therefore, we may now justify taking the $m \to 0$ limit while holding $\nu$ fixed; after which, if we then carry out the multiple derivatives with respect to $\widehat{n}\cdot\widehat{n}'$, we will be brought to
\begin{align}
\label{GreensFunction_dSphere_Massless}
G_d[\widehat{n}\cdot\widehat{n}']
= \frac{\Gamma[\frac{d-1}{2}]}{4 \pi^{\frac{d+1}{2}}} 
\int_0^\infty \dd r' \frac{r'^{\nu_\pm + \frac{d-3}{2}}}{\left( 1+r'^2-2r' \widehat{n}\cdot\widehat{n}' \right)^{\frac{d-1}{2}}}  .
\end{align}
This integral converges for all real non-zero $m$. It diverges if $m=0$ \cite{Chu:2013hra}; and also diverges if $m$ is purely imaginary\footnote{This is the case where the Helmholtz operator is the massless scalar wave operator in frequency space.} -- for $\nu_+ = \sqrt{(d-1)^2/4 + |mR|^2} > (d-1)/2 \geq 1/2$ we would have an IR divergence from the upper limit of integration $\int^\infty \dd r' r'^{\nu_+ - 1}$; for $\nu_- = -\sqrt{(d-1)^2/4 + |mR|^2} < -(d-1)/2$ we would have a UV divergence from the lower limit of integration $\int_0 \dd r' r'^{\nu_- + (d-3)/2}$. We also may compare eq. \eqref{GreensFunction_dSphere_Massless} with eq. (99) of \cite{Chu:2013hra} and note that we could have arrived at the former by using the Green's function of the Laplacian (i.e., with $m=0$) in Euclidean space and simply replaced the $r'^{d-2}$ in eq. (99) of \cite{Chu:2013hra} with $r'^{\nu_\pm + (d-3)/2}$.

{\bf Result} \qquad Nevertheless, if we temporarily assume that $m$ is real and refer to 8.714.2 of \cite{G&S}, we will in fact arrive at the final result for the Green's function of the Helmholtz operator $-\Box + m^2$ on the $d$-sphere:
\begin{align}
\label{GreensFunction_MassiveLaplacian}
G_d[\widehat{n}\cdot\widehat{n}';R]
= \frac{\Gamma\left[\frac{d-1}{2}-\nu\right] \Gamma\left[\frac{d-1}{2}+\nu \right]}{2^{\frac{d}{2}+1} \pi^{d/2} R^{d-2}}
\frac{P_{\nu-\frac{1}{2}}^{1-\frac{d}{2}}\left[-\widehat{n}\cdot\widehat{n}'\right]}{\left(1-\left(\widehat{n}\cdot\widehat{n}'\right)^2\right)^{\frac{d}{4}-\frac{1}{2}}},
\end{align}
with
\begin{align}
\nu \equiv \sqrt{\left(\frac{d-1}{2}\right)^2-(mR)^2}.
\end{align}
We have re-instated the radius $R$ of the $d$-sphere. Here, the $P_\nu^\mu[z]$ is the associated Legendre function of the first kind. At this point, we can check directly, using the ordinary differential equation obeyed by $P_\nu^\mu[z]$ (see 8.700.1 of \cite{G&S}), that $G_d$ in eq. \eqref{GreensFunction_MassiveLaplacian} is annihilated by $-\Box+m^2$ almost everywhere -- without any restriction on $d$ or $mR$.\footnote{For the reader's convenience, we record that $-\Box+m^2$ acting on a bi-scalar that depends on $\vec{\theta}$ and $\vec{\theta}'$ only through the dot product $\widehat{n} \cdot \widehat{n}'$ is
\begin{align}
\label{GreensFunction_dSphere_ODE}
R^2 (-\Box+m^2)G_d[\cos\theta\equiv\widehat{n}\cdot\widehat{n}']
=- \frac{\partial_\theta\left( \sin^{d-1}\theta \partial_\theta G_d[\cos\theta]\right)}{\sin^{d-1}\theta} + (mR)^2 G_d[\cos\theta] .
\end{align}} 
By re-expressing the associated Legendre function in terms of the hypergeometric function $\,_2F_1$ (see 8.704 of \cite{G&S}), followed by transforming $\,_2F_1$ appropriately to study its $\widehat{n}\cdot\widehat{n}' \to \pm 1^\mp$ limits, one may show that the Green's function in eq. \eqref{GreensFunction_MassiveLaplacian} is singular when and only when the observer lies on top of the source, i.e., when $\widehat{n} \cdot \widehat{n}' \to 1^-$. Also note that $G_d[\widehat{n}\cdot\widehat{n}';R]$ is in fact invariant under the replacement $\nu \to -\nu$ because $P_{-\nu-1}^\mu[z]=P_\nu^\mu[z]$; we may recognize the two possible choices of signs $\pm \nu$ as arising from the $I_{\nu_\pm}[mr']/(m r')^{\frac{d-1}{2}}$ which encoded the two possible charge densities we began with in equations \eqref{GreensFunction_dSphere_Even_Integral} and \eqref{GreensFunction_dSphere_Odd_Integral}.\footnote{A previous derivation of eq. \eqref{GreensFunction_MassiveLaplacian} can be found, for instance, in \cite{Szmytkowski}. The $N$ there is $d$ here; $\lambda(\lambda+N-1)$ there is $-m^2$ here; and to relate the Gegenbauer to the associated Legendre function, consult 8.936.1 of \cite{G&S}.}

We observe that $m$ occurs in eq. \eqref{GreensFunction_MassiveLaplacian} only through the combination $mR$. As one takes the massless limit, this amounts to taking the size of the $d$-sphere to zero: $R \ll m^{-1}$. (The result in eq. \eqref{GreensFunction_MassiveLaplacian} diverges in the $m\to 0$ limit because $\Gamma[0]$ is singular.) In this sense the precisely $m=0$ situation is disconnected from the finite $m$ setup. This is consistent with the fact that the minimum number of allowed sources in the former is two (one positive and one negative) as opposed to one; see \cite{Chu:2013hra} for a discussion  on the topological obstructions for the massless case.

The reader may complain at this point that the embedding method has forced us to undergo quite a few mathematically dubious steps in order to obtain the final result in eq. \eqref{GreensFunction_MassiveLaplacian}. This may be contrasted against the alternate method of solving the ordinary differential equation (see eq. \eqref{GreensFunction_dSphere_ODE}) $(-\Box+m^2)G_d[\cos\theta\equiv\widehat{n}\cdot\widehat{n}'] = 0$ (almost everywhere), whose general solution would be a superposition of the right hand side of eq. \eqref{GreensFunction_MassiveLaplacian} and the same expression with $P_{\nu-\frac{1}{2}}^{1-\frac{d}{2}}\left[-\widehat{n}\cdot\widehat{n}'\right]$ replaced with $Q_{\nu-\frac{1}{2}}^{1-\frac{d}{2}}\left[-\widehat{n}\cdot\widehat{n}'\right]$. However, one would then have to argue that the term containing $Q_{\nu-\frac{1}{2}}^{1-\frac{d}{2}}\left[-\widehat{n}\cdot\widehat{n}'\right]$ necessarily blows up at the antipodal point $\widehat{n} \cdot \widehat{n}' = -1$ and thereby violates the boundary condition that there should be one and only one point source on the $d$-sphere. Within the embedding method, on the other hand, the boundary condition is built in from the outset: that there is exactly one source on the $d$-sphere is because, from the perspective of the ambient Euclidean space experimentalist, the line source in eq. \eqref{GreensFunction_dSphere_LineSource} pierces the $d$-sphere at only one point.

{\bf Conformal Flatness} \qquad Like de Sitter spacetime, the geometry of the $d$-sphere is conformally flat. Specifically, redefining the $d$th angular coordinate via
\begin{align}
\theta \equiv 2 \tan^{-1}[\chi]
\end{align}
would bring the metric on the sphere to the form
\begin{align*}
\dd\Omega_d^2 = \left(\frac{2}{1+\chi^2}\right)^2 \left(\dd\chi^2 + \chi^2 \dd\Omega_{d-1}^2\right), \qquad \chi \geq 0 .
\end{align*}
We may thus perform another check on our Green's function result in eq. \eqref{GreensFunction_MassiveLaplacian}, by recovering the $d$-dimensional Euclidean Green's function of the Laplacian when we set
\begin{align*}
(mR)^2 \to \frac{d(d-2)}{4}, \qquad \nu \to \frac{1}{2} .
\end{align*}
Let us exploit the spherical symmetry of the problem and place the source at the north pole, so that $\chi'=0$ and $\widehat{n}\cdot\widehat{n}' = \cos\theta$. 
Upon glancing at eq. (47) of \cite{Chu:2013hra}, we expect to obtain, when $\nu=1/2$,
\begin{align}
G_d[\widehat{n}\cdot\widehat{n}';R] = \frac{1}{\Omega^{\frac{d}{2}-1}[\chi] \Omega^{\frac{d}{2}-1}[\chi']} \frac{\Gamma\left[\frac{d}{2}-1\right]}{4\pi^{d/2}\chi^{d-2}}, \qquad \qquad
\Omega[\chi] = \frac{2}{1+\chi^2}, \ \Omega[\chi'] = 2 .
\end{align}
This is confirmed once we register the identity 
\begin{align*}
P_0^\mu[\cos\theta]
= \frac{\cot^\mu\left[\theta/2\right]}{\Gamma[1-\mu]},
\qquad 0 < \theta < \pi .
\end{align*}
Notice our Green's function written in conformally flat coordinates appears incorrect at $d=2$. In two dimensions, a minimally coupled massless scalar is already conformally invariant; see eq. \eqref{ConformallyInvariantAction}. There, one would expect,
\begin{align}
\label{GreensFunction_dSphere_2D_WrongIofII}
G_2[\widehat{n}\cdot\widehat{n}';R] = -\frac{\ln\chi}{2\pi} + \text{constant}
\end{align}
i.e., without the conformal factors. Yet, performing dimensional regularization -- expanding our result here about $d=2$ -- would give an answer that depends on $\Omega[\chi]$ and $\Omega[\chi']$:
\begin{align}
\label{GreensFunction_dSphere_2D_WrongIIofII}
G_2[\widehat{n}\cdot\widehat{n}';R] = -\frac{\ln[\sqrt{\Omega[\chi] \Omega[\chi']} \cdot \chi]}{2\pi} + \text{constant}
\end{align}
The reason is that, since the conformal limit in two dimensions coincides with the massless limit, topology informs us there is no solution describing the field of a single point charge on a 2-sphere, so neither eq. \eqref{GreensFunction_dSphere_2D_WrongIofII} nor eq. \eqref{GreensFunction_dSphere_2D_WrongIIofII} are legitimate answers.

\section{Conclusions}
\label{Section_End}

We have succeeded, in this paper, to extend the formalism in \cite{Chu:2013hra} to compute the Green's function of the operator $\Box + m^2$, for $m \neq 0$, by using the flat space(time) Green's function in which the curved space(time) is embedded. The specific type of embedding where our general formula in eq. \eqref{GreensFunction_GeneralFormula} is applicable can be found in eq. \eqref{Metric_GeneralEmbeddingForm}. The result when $\Box$ is with respect to the $d$-dimensional de Sitter metric can be found in equations \eqref{GreensFunction_deSitter_Closed} through \eqref{GreensFunction_deSitter_Odd}. For $\Box$ with respect to the (negative of) the metric on the $d$-sphere, the result is eq. \eqref{GreensFunction_MassiveLaplacian}. For de Sitter spacetime the embedding method has, once again, not only allowed us to directly extract (and separate) the null cone versus tail pieces of the massive scalar Green's function in arbitrary dimensions $d \geq 2$, it has also taught us that we may identify separate sources for them. (See Fig. \eqref{Figure_CausalStructureFromEmbedding}.) Moreover, we have clarified that there are two distinct scalar charge densities, associated with the two independent solutions of the eigenvector equation \eqref{EigenvectorEquation}, that can act as the ambient Minkowski source for the de Sitter Green's function. On the other hand, the integrals encountered in the $d$-sphere case were significantly more nuanced than for its de Sitter counterpart; nonetheless, we have argued that, even though the intermediate steps involved assuming $mR$ and $d$ to lie in some restricted range, these assumptions may be dropped once the final result was obtained.\footnote{In this paper we are primarily interested in the causal structure of physical signals of classical scalar field theories in curved space(time)s embeddable in some ambient flat space(time) via eq. \eqref{Metric_GeneralEmbeddingForm}. An analogous application to quantum field theory can be found in \cite{Bertola:2000mx}.}

We end by highlighting that our investigations have uncovered -- see equations \eqref{GreensFunction_deSitter_Even_Integral_III}, \eqref{GreensFunction_deSitter_Odd_Integral_III} and \eqref{GreensFunction_dSphere_Massless} -- that the Minkowski/Euclidean experimentalists could also use massless scalar fields to achieve the same goal of sourcing the massive scalar Green's function on the curved sub-manifold of the observer's world. This suggests the possibility that, even within the class of embeddings we are considering here, there could be alternate formalisms that are available for understanding curved space(time) Green's functions. This in turn gives us hope that, perhaps, a more general formalism may be found that could then be extended and applied to cosmological and black hole spacetimes.

\section{Acknowledgments}

I wish to thank the anonymous referee for providing references \cite{Szmytkowski} and \cite{Rosenthal:2003qr}. We have employed \mma \ \cite{Mathematica} for much of the analytic calculations in this paper. This work was supported by NSF PHY-1145525 and funds from the University of Pennsylvania.

\end{document}